\documentclass[aip,amsmath,amssymb,jcp,preprint
 ,12pt
]{revtex4-1}

\usepackage{amsfonts}
\usepackage{amsmath}
\usepackage{amssymb}
\usepackage{amsthm}
\usepackage{appendix}
\usepackage{datetime}
\usepackage{dcolumn} 
\usepackage{enumerate}
\usepackage[T1]{fontenc}
\usepackage{epigraph}
\usepackage{graphicx} 
\usepackage[utf8]{inputenc}
\usepackage{hyperref}
\hypersetup{
    colorlinks=true,
    linkcolor=blue,
    filecolor=magenta,
    urlcolor=cyan,
}
\usepackage{mathptmx}
\usepackage{soul} 
\usepackage{ulem} 
\usepackage{xcolor}

\newcommand{\msout}[1]{\text{\sout{\ensuremath{#1}}}}

\newcommand{\R}{\mathbb{R}}  

\newcommand{\bfr}{{\bf r}}
\newcommand{\bfs}{{\bf s}}

\newcommand{\bfR}{{\bf R}}

\newcommand{\expect}[2]{\ensuremath{\langle #2 \vert #1 \vert #2 \rangle}}

\DeclareMathOperator{\dd}{d}
\DeclareMathOperator{\erf}{erf}
\DeclareMathOperator{\erfc}{erfc}
\makeatletter
\newcommand{\thickbar}{\mathpalette\@thickbar}
\newcommand{\@thickbar}[2]{{#1\mkern1.5mu\vbox{
  \sbox\z@{$#1\mkern-1.5mu#2\mkern-1.5mu$}%
  \sbox\tw@{$#1\overline{#2}$}%
  \dimen@=\dimexpr\ht\tw@-\ht\z@-.8\p@\relax
  \hrule\@height.8\p@ 
  \vskip\dimen@
  \box\z@}\mkern1.5mu}
}
\makeatother
\renewcommand*{\bar}{\thickbar}

\newcommand{\updated}[2]{{#2}}        

\begin{document}

\title{\updated{Correction of the basis set error due to the absence of the electron-electron cusp in the wave function by using an adiabatic correction}{Correcting basis-set errors due to the absence of the electron-electron cusp using an adiabatic correction}}

\author{Anthony Scemama}
\affiliation{Laboratoire de Chimie et Physique Quantiques (UMR 5626),
Université de Toulouse, CNRS, UPS, France}
\email{scemama@irsamc.ups-tlse.fr}

\author{Andreas Savin}
\affiliation{Laboratoire de Chimie Th\'eorique, CNRS and Sorbonne University \\ 4 place Jussieu, 75252 Paris, France}
\email{andreas.savin@lct.jussieu.fr}

\begin{abstract}
This article proposes an analytical method to address the slow convergence of electronic structure calculations caused by the inability of finite one-particle basis sets to describe the electron-electron cusp.
  An equivalence is \updated{made}{established} between a calculation using a finite basis set with the physical Coulomb interaction and a calculation using a complete basis set with a model interaction (specifically, the error-function screened Coulomb potential characterized by a range-separation parameter $\mu$). By leveraging the adiabatic connection formalism, a simple, parameter-free correction formula is derived.
It depends only on the on-top pair density and a locally defined range-separation parameter ($\mu$) derived from the basis set itself.
This `adiabatic connection based basis-set error correction' (ABC) is derived from the asymptotic expansion of the wave function at large $\mu$ for small inter-electronic distances.
Therefore it is applicable to both ground and excited states without the restriction imposed by the Hohenberg-Kohn theorem.
Numerical tests illustrate that the method achieves chemical accuracy using smaller basis sets than typically required.

\end{abstract}


\maketitle

\makeatletter
\def\l@section#1#2{\@dottedtocline{1}{0.5em}{1.8em}{#1}{#2}}
\def\l@subsection#1#2{\@dottedtocline{2}{2.0em}{1.8em}{#1}{#2}}
\makeatother

\centerline{\today \hspace{2pt} at \currenttime}


\section{Introduction}

One-particle basis functions are omnipresent in electronic structure calculations.
They are also used to construct Slater determinants as $N$-body basis functions.
\updated{One has to be careful to include important basis functions
such as those}{Care must be taken to include basis functions} needed to describe polarization~\cite{RooSad-CP-85} (the
effect of an electric field, external to the system or due to other
nuclei in the system), to add diffuse functions to describe excited (Rydberg) states~\cite{KauBauJun-JPB-89}, negative ions~\cite{KenDunHar-JCP-92}, or some needed to describe near-degeneracy (orbitals not present in the reference determinant such as $p$ functions for the Be atom~\cite{BarLinShu-65}).
However, it has been recognized that the convergence of the Slater determinant expansion is very slow due to \updated{}{the} singularity of the Coulomb interaction, when two electrons occupy the same position.~\cite{KutMor-JCP-92}
\updated{To treat it, several ways have been proposed}{Several approaches have been proposed to address this problem}: extrapolation to the complete basis set~\cite{HelKloKocNog-97,HalHelJor-98}, or explicitly correlated methods, such as F12~\cite{Ten-TCA-12,TewKlo-MP-10}.
More recently, it has been proposed to use density functionals \updated{to}{for} the same purpose.~\cite{GinPraFerAssSavTou-JCP-18}
In a variant of this method, one also \updated{uses}{includes} a dependence on the on-top
pair density.~\cite{FerGinTou-JCP-19,GinSceLooTou-JCP-20}
\updated{The objective of this paper is to consider the line of thought that was at the origin of the latter approach}{This paper follows the line of thought that motivated the latter approach},~\cite{GorSav-PRA-06} without employing a density functional approach.
\updated{This way, there is no need to use the Hohenberg-Kohn theorem}{In this way, there is no need to invoke the Hohenberg-Kohn theorem} (and thus the restriction to the ground state), and \updated{making extensions to properties is trivial}{extensions to properties are straightforward} (via the Hellmann-Feynman theorem).
\updated{}{The method proposed in this paper falls in this class of methods, trying to correct for the error made by basis sets in treating the domain where the distance between a pair of electrons is small, and not the other limitations of the basis sets.
On the same topic, one can point the reader to ref.~\onlinecite{HapTucModGolVeiPer-JPCL-25} that also aims to correct the basis set errors due to the existence of the electron-electron cusp, without relying on density functionals,
but by introducing an effective Hamiltonian.}

\updated{In this paper, we consider full configuration interaction (FCI)
energies as estimated from selected configuration interaction~\cite{HurMalRan-73} calculations
and correct for the missing correlation energy by using an expression of the form derived from a Taylor expansion for large $\mu$, starting with $-|a_3| \mu^{-3}$, where $\mu$ is a parameter characterizing the range of the model interaction (with the dimension of an inverse length).
It connects the finite basis set calculation to a complete basis set representation.
Of course, this expression cannot be used for small $\mu$.
To correct for this, ref.~\onlinecite{GinSceLooTou-JCP-20} uses a regularization in a form
}{}
\updated{By using for $e_c$ the Perdew-Burke-Ernzerhof (PBE) density functional~\cite{PerBurErn-PRL-96}, it recovers PBE at $\mu=0$.
In this paper, we do not use any regularization, and compare our
results with those obtained using the ansatz in eq. \dots
as well as those obtained by complete basis set two-point extrapolation using the cardinal number characterizing the basis set (eq. (7) of ref.~\onlinecite{HelKloKocNog-97}).}

To facilitate the reading of the paper, \updated{we recall in the following
sections the foundation of our method~\cite{Sav-JCP-20,SavKar-JPCA-23,SceSav-JCC-24,SceSav-Gus-JPCA-24,SceSav-Trygve-JPCA-24}}{we review the foundations of our method~\cite{Sav-JCP-20,SavKar-JPCA-23,SceSav-JCC-24,SceSav-Gus-JPCA-24,SceSav-Trygve-JPCA-24} in the following sections}, namely considering the short-range behavior of the wave function for model potentials without \updated{}{a} singularity, and the adiabatic connection for using it to obtain the energy of the physical system (with Coulomb interaction between electrons) providing an analytic expression.
In the \updated{following}{next} section, we show how a finite basis set calculation can be connected to \updated{a complete basis set one}{a complete basis-set calculation} using a model potential. The correction to the basis-set error is made by considering the transformation from this model to the physical system.
We use the acronym ABC for the method proposed in this paper: ``Adiabatic connection based Basis-set error Correction''.
Next, we give some numerical examples.
Finally, we point out some of the limitations of our implementation, and \updated{paths to improve it}{possible paths for improvement}.

\section{Foundation of the method}

\subsection{Short range behavior of the wave function}

\updated{When one considers two electrons, at positions $\bfr_1$, $\bfr_2$, and distance $r=|\bfr_1-\bfr_2|$}{Consider two electrons at positions  $\bfr_1$, $\bfr_2$, separated by $r=|\bfr_1-\bfr_2|$}, the coalescence condition dictates that the wave function behaves for $r \rightarrow 0$ as~\cite{Kat-CPAM-57, FouHofHofOst-CMP-09, KurNakNak-AQC-16}
\begin{equation}
 \Psi(\bfr_1,\bfr_2,\dots) = \sum_{\ell,m} c_{\ell,m}(\bfR,\dots)  \psi_\ell(r) Y_{\ell,m}(\Omega_\bfr) +\dots ,
 \label{eq:psi-kato}
\end{equation}
where $\bfR$ is a coalescence point, $Y_{\ell,m}$ are the spherical
harmonics, $\Omega_\bfr$ \updated{is the solid angle related to the orientation of}
{denotes the angular coordinates of}
$\bfr=\bfr_1-\bfr_2$, and
\begin{equation}
\label{eq:psi-l}
 \psi_\ell(r) = r^\ell \left(1+\frac{1}{2\ell+2} r \right)  .
\end{equation}
\updated{Through}{Because of} the factor $r^{\ell}$ present in $\psi_\ell$, the electrons are kept further apart with increasing $\ell$.

For expectation values of short-ranged two-particle operators (that depend on $r$, and possibly on $\bfR$), we use the spherically averaged pair density that, for $r \rightarrow 0$, is
\begin{equation}
\label{eq:p2}
 P_2(\bfr_1,\bfr_2) = \sum_\ell C_\ell(\bfR) r^{2 \ell} \left(1+\frac{1}{\ell+1} r \right) +\dots
\end{equation}
In particular, at $r=0$, we obtain for $C_\ell(\bfR)$
i) for $\ell=0$, the on-top pair density, $P_2(\bfr_1=\bfR,\bfr_2=\bfR)$, while
ii) for $\ell>0$, $\lim_{\bfr_1 \rightarrow \bfr_2} P_2(\bfr_1,\bfr_2)/r^{2 \ell}$, that \updated{can be also expressed by}{can also be expressed in terms of} derivatives at the coalescence point.
Only the $\ell=0$ component, $\psi_0$, contributes to a non-vanishing on-top pair density.
The value of $C_\ell(\bfR)$ is not determined by \updated{short-range only}{the short-range behavior alone}: \updated{one can rather see it as given by a ``normalization''}{rather, it can be viewed as a normalization factor}, that is, determined by the overall behavior of the wave function.

\subsection{Adiabatic connection}

We \updated{deviate}{depart} from the physical Hamiltonian by changing only the interaction between electrons to
\begin{equation}
 \label{eq:wlam}
 w(r,\lambda,\mu) = w(r,\mu) + \lambda \bar{w}(r,\mu).
\end{equation}
For $\lambda=0$, we have a model interaction,
\begin{equation}
\label{eq:w}
 w(r,\mu) = \frac{\erf(\mu r)}{r} ,
\end{equation}
while for $\lambda=1$ we have the physical Coulomb interaction.
We denote the interaction missing in the model by
\begin{equation}
\label{eq:wbar}
 \bar{w}(r,\mu) = \frac{\erfc(\mu r)}{r} = \frac{1}{r} - \frac{\erf(\mu r)}{r} ;
\end{equation}
\begin{equation}
 \label{eq:Wbar}
 \bar{W}(\mu) = \sum_{i,j>i} \bar{w}(|\bfr_i-\bfr_j|,\mu).
\end{equation}
$H(\lambda,\mu)$ is the Hamiltonian \updated{where only the interaction $w(r,\lambda,\mu)$, eq.~\eqref{eq:wlam} is different from the physical Coulomb one}
{in which only the interaction $w(r,\lambda,\mu)$, eq.~\eqref{eq:wlam} differs from the physical Coulomb interaction}.
$\Psi(\lambda,\mu)$ is an eigenfunction of $H(\lambda,\mu)$.
To simplify the notation, we do not specify the eigenfunction by an index, although we keep in mind that $\Psi(\lambda,\mu)$ \updated{does not have to}{need not} be the ground state. This flexibility is a major advantage of ABC, as it allows for the treatment of excited states without the inherent ground-state restriction of the Hohenberg-Kohn theorem.

When using $H(\lambda, \mu)$, all energy terms change with $\lambda$ and $\mu$.
However, one can correct $\expect{H}{\Psi(0,\mu)}$ and use only $\bar{W}$ by exploiting the adiabatic connection~\cite{Sav-JCP-20}.
Here, $H$ is the physical Hamiltonian and the total energy of the system is
\begin{equation}
 \label{eq:ac}
 E = \expect{H}{\Psi(0,\mu)}  + \int_0^1 \dd \lambda \, \left( \expect{\bar{W}(\mu)}{\Psi(\lambda,\mu)} - \expect{\bar{W}(\mu)}{\Psi(0,\mu)} \right)
\end{equation}

A key observation needed \updated{for obtaining}{to obtain} ABC is to realize that $\bar{w}$ is, for a sufficiently large value of $\mu$, a short-range operator and thus only the short-range behavior of the wave function is needed to evaluate the correction to $\expect{H}{\Psi(0,\mu)}$.
The form of $\Psi(\lambda,\mu)$ is similar to that of $\Psi$, eq.~\eqref{eq:psi-kato}.
The behavior of $\Psi(\lambda,\mu)$ for $r \rightarrow 0$ is known~\cite{SilUgaBoy-The-00, SavKar-JPCA-23}: $\psi_\ell(r)$, eq.~\eqref{eq:psi-l} is changed to $\psi_\ell(r,\lambda,\mu)$.
In particular,
\begin{align}                                                                                                                         \label{eq:psi-0-lambda-mu}
 \psi_0(r,\lambda,\mu) & =1 + \frac{1}{2}r +(1-\lambda ) \left(\left(\frac{1}{4 \mu ^2 r}+\frac{r}{2}\right) \erf(\mu  r)+\frac{e^{- \mu ^2 r^2}}{2
   \sqrt{\pi } \mu }-\frac{r}{2}\right)
 \\
 \label{eq:psi-1-lambda-mu}
 \psi_1(r,\lambda,\mu) & = r \left( 1 + \frac{1}{4} r + (1-\lambda ) \left(\left(\frac{1}{16 \mu ^4 r^3}+\frac{r}{4}\right) \erf(\mu  r)+e^{- \mu ^2 r^2}
   \left(\frac{1}{4 \sqrt{\pi } \mu }-\frac{1}{8 \sqrt{\pi } \mu ^3 r^2}\right)-\frac{r}{4}\right) \right) .
\end{align}
In the limits $\lambda=1$ or $\mu\to\infty$, where the model interaction reverts to the physical Coulomb potential, the functions
$\psi_{\ell}(r,\lambda,\mu)$ reduce to the standard $\psi_{\ell}(r)$ form defined in eq.~\eqref{eq:psi-l}.
To leading order in $1/\mu$, we have~\cite{SavKar-JPCA-23}:
\begin{align}
 \label{eq:ac-0-corr}
 \int_0^1 \dd \lambda \, \int_0^\infty \dd r \, \left(|\psi_0(r,\lambda,\mu)|^2 - |\psi_0(r,0,\mu)|^2 \right) \frac{\erfc(\mu r)}{r} 4 \pi r^2 & = - \frac{4 (\sqrt{2} -1)\sqrt{\pi}}{3} \mu^{-3} \\
 \label{eq:ac-1-corr}
 \int_0^1 \dd \lambda \, \int_0^\infty \dd r \, \left(|\psi_1(r,\lambda,\mu)|^2 - |\psi_1(r,0,\mu)|^2 \right) \frac{\erfc(\mu r)}{r} 4 \pi r^2 & = - \frac{(3\sqrt{2} -4)\sqrt{\pi}}{5} \mu^{-5}
\end{align}

\updated{}{In general, as $\psi_\ell$ behaves as $r^\ell$ for $r \rightarrow 0$,  we find by the substitution $x=\mu r$, that these integrals decay as $\mu^{-(2 \ell +1)}$.

When considering the short-range behavior of $\Psi(\lambda,\mu)$ one obtains that the terms depending on $\bfR$ do not depend on $\lambda$ or $\mu$.
This can be understood by the fact that the expressions were derived by perturbing the exact solution of the Schr\"odinger equation to treat the model system.

Now let us go back to the adiabatic connection, eq.~\eqref{eq:ac}.
Only the expectation values $\expect{\bar{W}}{\Psi(\lambda,\mu)}$ are needed (and not those of other operators).
As, for large $\mu$, the operator $\bar{W}$ is short-ranged and local, one needs only the short-range part of the pair density generated by $\Psi(\lambda,\mu)$, $P_2(\bfr_1,\bfr_2;\lambda,\mu)$:
\begin{align}
 \expect{\bar{W}}{\Psi(\lambda,\mu)} & = \frac{1}{2} \int_{\R^3} \dd \bfr_1 \int_{\R^3} \dd \bfr_2 P_2(\bfr_1,\bfr_2;\lambda,\mu) \bar{w}(r_{12})\\
 & = \frac{1}{2} \sum_\ell \int_{\R^3} \dd \bfR \; C_\ell(\bfR) \int_0^\infty \dd r \, 4 \pi r^2  |\psi_\ell(r,\lambda,\mu)|^2 + \dots
\end{align}
We have seen that the integrals over $r$ decay as $\mu^{-(2 \ell + 1)}$.
The dominating term originates from $\ell=0$, eq.~\eqref{eq:ac-0-corr}, and is in $\mu^{-3}$.
Using the fact that $C_0(\bfR)$ is just the on-top pair density, $P_2(\bfR,\bfR)$, eq.~\eqref{eq:p2}, the leading correction to $\expect{H}{\Psi(0,\mu)}$ is coming from the $\ell=0$ component,
\begin{align}
  \int_0^1 \dd \lambda \, \left(
    \expect{\bar{W}(\mu)}{\Psi(\lambda,\mu)}  -
    \expect{\bar{W}(\mu)}{\Psi(0,\mu)} \right) = \nonumber \\
   -\frac{1}{2} \int_{\R^3} \dd \bfR \int \dd P_2(\bfR,\bfR) \, \frac{4 (\sqrt{2} -1)\sqrt{\pi}}{3} \mu^{-3} +
    \dots
 \label{eq:mu-correction}
\end{align}
Further details can be found in refs.~\onlinecite{Sav-JCP-20} or ~\onlinecite{SavKar-JPCA-23}.
}

Eq.~\eqref{eq:mu-correction} reveals the appearance of the exact pair density $P_2$, normalized to $N(N-1)$, at the coalescence point, $\bfR$.
This quantity is not assumed available, only the pair density for the
calculation with the model Hamiltonian, with the interaction $w(r,\mu)$, $P_2(\bfR,\bfR,\mu)$, is.
However the two can be related (starting from eq.~\eqref{eq:psi-0-lambda-mu}, for $r \rightarrow 0$ one obtains eq.~(30) in ref.~\onlinecite{GorSav-PRA-06}),
\begin{equation}
 \label{eq:p2-mu}
 P_2(\bfR,\bfR) = P_2(\bfR,\bfR,\mu) \left( 1 + \frac{2}{\sqrt{\pi} \mu} \right)^{-1} + \dots
\end{equation}
So, once we have solved the model problem at a chosen value of $\mu$, we can obtain the correction using only information gained from this calculation,
\begin{equation}
 \label{eq:e-approx}
  E = \expect{H}{\Psi(0,\mu)} -  \frac{1}{2} \int_{\R^3} \dd \bfR \, P_2(\bfR,\bfR,\mu ) \left( 1 + \frac{2}{\sqrt{\pi} \mu} \right)^{-1}  \ \frac{4 (\sqrt{2} -1)\sqrt{\pi}}{3} \mu^{-3} + \dots
\end{equation}

When $\mu$ is a given constant, only the system average of the on-top pair density,
\[ \int_{\R^3} \dd \bfR \int \dd P_2(\bfR,\bfR,\mu) \]
is needed.
One can also use the expectation value of some short-range operators such as $\bar{W}(\mu)$ to get an estimate of this term.~\cite{SceSav-Trygve-JPCA-24}

\updated{In the above derivation, there is no restriction to $\mu$ being a constant.}{
Up to now, $\mu$ was treated as a constant.
However, one can see already from its dimension of inverse length that a ``large'' value of $\mu$ has different meanings in regions where electrons are far apart or confined to a smaller region.~\cite{SavFla-IJQC-95}

In such a case the model error, that is, the difference between the expectation value of the Hamiltonian with the exact and the model wave function is given to leading order in $1/\mu$ by}
\begin{equation}
 \label{eq:correct-lbs}
  \updated{}{
   \Delta E_\text{ME} (\mu)
 }
  = \updated{}{- \frac{2 (\sqrt{2} -1)\sqrt{\pi}}{3}} \int_{\R^3} \dd \bfR  \; P_2(\bfR,\bfR,\mu(\bfR)) \left( 1 + \frac{2}{\sqrt{\pi} \mu(\bfR)} \right)^{-1}   \mu(\bfR)^{-3}
\end{equation}
has to be computed for the system of interest.

When $\mu$ is large, eq.~\eqref{eq:e-approx} reflects the exact behavior.
This is convenient when the model calculation is cheaper than the physical one, and \updated{it is expected to be so}{this is expected to be the case}, because (in contrast to the Coulomb interaction) there is no singularity when $r \rightarrow 0$.
However, the gain decreases as the model interaction approaches the physical interaction (as eq.~\eqref{eq:e-approx} becomes more accurate).
If we can extend the approximation to perform better for weaker interactions (smaller $\mu$) \updated{we can gain by using smaller basis sets}{one can benefit from using smaller basis sets}.
Extensions of eq.~\eqref{eq:e-approx} are needed, because eq.~\eqref{eq:e-approx} was derived from an expansion in $1/\mu$ retaining only the leading term.
One could add more terms~\cite{KarSav-MP-22}, regularize (taking advantage of some knowledge of the models at small $\mu$), extracting more information from other properties of the system, etc.
Instead of extending the domain of validity of eq.~\eqref{eq:e-approx} to smaller $\mu$, \updated{we recognize that our objective is}{we aim in this paper} to work with a small basis set.
While the preceding derivation assumes a complete basis set for the model interaction, the following section establishes an equivalence between model potentials \updated{}{(not necessarily of the form eq.~\eqref{eq:w})} and the use of finite basis sets, providing a practical framework for basis-set error correction.

\section{ Correcting the basis-set error}

\subsection{Approximation for the projection operator onto a basis set}

We aim to establish a link between a calculation using a finite basis set $B$ and a model interaction represented in a complete basis set.
In this paper, \updated{we argue}{we make the argument} using a rough approximation of the projector onto the basis set, $\mathcal{P}_B(\bfr_i,\bfr'_i) $ that satisfies
\begin{equation}
 \int \dd \bfr'_i \mathcal{P}_B(\bfr_i,\bfr'_i) \varphi (\bfr'_i) =
 \begin{cases}
  \varphi(\bfr_i) &  \varphi \in B \\
  0 & \varphi \notin B
 \end{cases}
 \label{eq:proj-def}
\end{equation}
for some one-particle function $\varphi(\bfr_i)$, $\bfr_i \in \R^3$.
\updated{It can be an orbital, but also}{It may be either an orbital or} an orbital product.
We can write the projector in a finite basis set of non-orthogonal basis functions, $\varphi_j$, that span $B$  as
\begin{equation}
 \mathcal{P}_B(\bfr_i,\bfr'_i) = \sum_{j,k} \varphi_j(\bfr_i) S_{jk}^{-1} \varphi_k^*(\bfr_i') ,
\label{eq:proj-fin}
\end{equation}
where $S^{-1}$ is the inverse of the overlap matrix between the $\varphi_j \in B$.
\updated{We consider here that the basis set is local, like a Gaussian basis set.}{}

In a complete basis set, $\mathcal{P}_B(\bfr_i,\bfr_i')  \rightarrow \mathcal{P}(\bfr_i,\bfr'_i)$,
\begin{equation}
 \mathcal{P}(\bfr_i,\bfr'_i) = \delta(\bfr_i-\bfr_i')  ,
\label{eq:proj-cbs}
\end{equation}
with $i=1,2,\dots$
For practical purposes, let us approximate the projector on a basis set,  $\mathcal{P}_B$, by spreading $\delta(\bfr_i-\bfr_i')$ using a Gaussian
\begin{equation}
 \mathcal{P}_B(\bfr_i,\bfr'_i) \approx \mathcal{P}_\epsilon(\bfr_i,\bfr_i') = (2 \pi \epsilon)^{-3/2} e^{-\frac{1}{2 \epsilon} (\bfr_i-\bfr_i')^2} .
\label{eq:proj-eps}
\end{equation}
As $\mathcal{P}_\epsilon$ is not idempotent, it must be viewed as a rough approximation rather than a strict projection operator.
\updated{}{
  However, it allows us to easily start exploring the method we propose, and establishing a connection with density functional based approaches that were shown in recent years to provide good corrections to basis-set errors.~\cite{GinPraFerAssSavTou-JCP-18,GinSceTouLoo-JCP-19,LooPraSceTouGin-JPCL-19,GinSceLooTou-JCP-20,LooPraSceGinTou-JCTC-20,GinTraPraTou-JCP-21,TraTouGin-JCP-22,TraGinTou-JCP-23,TraTouGin-24}
}

To analyze the impact of the finite basis set on the description of
short-range electron interactions, we transform the two-particle
projection operator into center-of-mass and relative coordinates,
defined as $\bfr_{cm} = (\bfr_1 + \bfr_2)/2$ and $\bfr = \bfr_1 - \bfr_2$.
By applying the Gaussian product rule (see, \textit{e.g.}, eqs.~A.2-A.5
in ref.~\onlinecite{SzaOst-BOOK-89}.), the product of the approximate
single-particle projectors can be factored into a center-of-mass
component and a relative distance component:

\begin{equation}
P_\epsilon(\mathbf{r}_1, \mathbf{r}'_1) P_\epsilon(\mathbf{r}_2, \mathbf{r}'_2) = P_{\epsilon/2}(\mathbf{r}_{cm}, \mathbf{r}'_{cm}) P_{2\epsilon}(\mathbf{r}, \mathbf{r}').
\label{eq:proj-r}
\end{equation}

\subsection{The effect of the basis set on the wave function}

Let us consider the effect of the projection for small distances between electrons, that is we apply $\mathcal{P}_{2\epsilon}$ \updated{on}{to} $\psi_0$, eq.~\eqref{eq:psi-l},
\begin{align}
 \nonumber
 \int_{\R^3} \dd  \bfs\, \mathcal{P}_{2 \epsilon}\left(\bfr,\bfs\right)\left(1+\frac{1}{2}s\right) & =
 \int_0^\infty \dd s \, 2 \pi s^2 \int_{-1}^1 \dd t \, \left(4 \pi \epsilon\right)^{-3/2} e^{-\frac{1}{4 \epsilon}(r^2+s^2-2 r \, s \, t)} \left(1+\frac{1}{2}s\right) \\
 \label{eq:proj-kato}
 & = 1 + \left( \frac{1}{2} r  +\frac{\epsilon}{r} \right) \text{erf}\left( \frac{r}{\sqrt{4 \epsilon }} \right) +\sqrt{\frac{\epsilon}{\pi}} \,  e^{-\frac{r^2}{4 \epsilon }} .
\end{align}
The r.h.s. of the preceding equation recovers the wave function for a model interaction between electrons, $\psi_0(r,0,\mu)$, eq.~\ref{eq:psi-0-lambda-mu}, setting $\mu$ to
\begin{equation}
\label{eq:mu-eps}
 \mu_\epsilon=\frac{1}{\sqrt{4 \epsilon}}.
\end{equation}

Consequently, \updated{}{when using the Gaussian approximation for the projector,} the basis set yields a wave function that exhibits short-range deviations analogous to
those of the model wave function characterized by $\mu_\epsilon$.
We can thus correct the basis-set error in the \updated{way the model error was corrected}{same way as the model error} (described by eqs.~\eqref{eq:mu-correction} and \eqref{eq:p2-mu}, by setting $\mu:=\mu_\epsilon$).

Eq.~\eqref{eq:mu-eps} can alternatively be derived by projecting $1/r$ onto the basis.
For this, let us use the notations
\[ \mathcal{P}_B(i) = \mathcal{P}_B(\bfr_i,\bfr_i') \]
and
\[ \left( \varphi_j |w(r,\infty)| \varphi_k \right) = \int_{\R^3} \dd \bfr_1 \int_{\R^3} \dd \bfr_2 \, \varphi_j(\bfr_1) \frac{1}{|\bfr_1-\bfr_2|} \varphi_k(\bfr_2) , \]
where $\varphi_i$ are orbital products.
\begin{equation}
 \left( \varphi_j |w(r,\infty)| \varphi_k  \right) = \left( \mathcal{P}_B(1) \varphi_j  |w(r,\infty)| \mathcal{P}_B(2) \varphi_k  \right) = \left( \varphi_j |\mathcal{P}_B(1) w(r,\infty)\mathcal{P}_B(2)| \varphi_k \right).
\end{equation}
We see that, in a calculation in basis set $B$, the Coulomb operator is ``seen'', not as $1/r$, but as  $\mathcal{P}_B(1) w(r,\infty)\mathcal{P}_B(2)$.
We approximate this operator replacing $\mathcal{P}_B$ by $\mathcal{P}_\epsilon$ and obtain
\begin{equation}
\label{eq:w-proj}
 \int_{\R^3} \dd \bfs_1 \int_{\R^3} \dd \bfs_2 \mathcal{P}_\epsilon(\bfr_1,\bfs_1) w(|\bfs_1-\bfs_2|,\infty)\mathcal{P}_\epsilon(\bfs_2,\bfr_2) = w(r,\mu_\epsilon),
\end{equation}
a result consistent with the previous derivation of $\mu_\epsilon$.
(Eq.~\eqref{eq:w-proj} can be easily obtained by considering the electron repulsion integral for $s$-type Gaussians, see, \textit{e.g.}, ref.~\onlinecite{SzaOst-BOOK-89}, eq.~A.41.)

\updated{}{
\subsection{The ABC approach in short}
\label{subs:ABC-short}


The ``Adiabatic connection based Basis-set error Correction'' (ABC) is based upon the observation that a finite basis set for the physical Hamiltonian produces a wave function that behaves at short range as the eigenfunction of a model Hamiltonian in a complete basis set, eq.~\eqref{eq:proj-kato}.
The basis-set error due to the inability to reproduce the electron-electron cusp can thus be corrected by computing the error of a model calculation in a complete basis set.

In principle, one should use the exact projection operator onto a basis to produce the model system to be corrected.
Instead, we use in this paper a simplified scheme: the projection operator is replaced by a Gaussian, eq.~\eqref{eq:proj-eps}.
We are motivated to use this approximate projector because the it generates the short-range behavior of the wave function of the  Hamiltonian  having an interaction term of the form $\erf(\mu r)/r$. This type of Hamiltonian has been amply used in the past and the expression for the error due to the model can be obtained analytically to leading order in $1/\mu$.
It is given by $\Delta E_\text{ME}(\mu)$, eq.~\eqref{eq:correct-lbs}.
To obtain the basis-set error correction within this approximation of the projection operator, $\mu(\bfR)$ in eq.~\eqref{eq:correct-lbs} is replaced by some $\mu_\epsilon(\bfR)$, defined by the projector, eq.~\eqref{eq:mu-eps}.
The total energy, corrected for the basis-set error, is estimated as
\begin{align}
 \nonumber
 E & \approx \expect{H}{\Psi_B} + \Delta E_\text{ME}(\mu_\epsilon) \\
  \label{eq:bsc}
 & \approx \expect{H}{\Psi_B} - \frac{2 (\sqrt{2} -1)\sqrt{\pi}}{3}  \int_{\R^3} \dd \bfR \; P_2(\bfR,\bfR,\mu_\epsilon(\bfR)) \left( 1 + \frac{2}{\sqrt{\pi} \mu_\epsilon(\bfR)} \right)^{-1}   \mu_\epsilon(\bfR)^{-3}
\end{align}
where $\Psi_B$ is the wave function --- ideally obtained in a full configuration interaction (FCI) calculation in the basis $B$.
In this paper we replace FCI by a selected CI~\cite{HurMalRan-73} that we consider equivalent.
In practice, other approximations to FCI can be watchfully used such as coupled cluster calculations.
Note that the correction of the basis-set error consists of using an expression that contains a numerical integration on $\R^3$, similar to what is done in density functional calculations.
In our case, eq.~\eqref{eq:bsc}, the integrand does not require the knowledge of the electron density (as in density functional calculations), but that of the on-top pair density.
Furthermore, it requires the knowledge of $\mu_\epsilon(\bfR)$.
The choice we make in this paper for it is described below, in subsection~\ref{subs-def-muB}.
As this choice is made to allow a direct comparison with an existing method, PBE-OT, we briefly mention the connection between PBE-OT and ABC.
}

\subsection{A short description of the PBE-OT method}

\updated{}{Being valid only to leading order in $1/\mu$, eq.~\eqref{eq:bsc} cannot be used for small $\mu$.
It is possible to regularize the integrand in the basis-set correction imposing a density functional result at $\mu=0$:
\begin{equation}
 \label{eq:reg}
  \Delta E_\text{reg} =  \int_{\R^3} \dd \bfR \frac{e_c(\bfR, \rho)}{1 - \frac{e_c(\bfR, \rho)}{\frac{2 (\sqrt{2} -1)\sqrt{\pi}}{3} P_2(\bfR,\bfR,\mu_\epsilon(\bfR)) \left( 1 + \frac{2}{\sqrt{\pi} \mu_\epsilon(\bfR)} \right)^{-1}   \mu_\epsilon(\bfR)^{-3}}} .
\end{equation}
Here, $e_c$ is the correlation energy density in some density functional approximation.
For example, in ref.~\onlinecite{GinSceLooTou-JCP-20} the Perdew-Burke-Ernzerhof (PBE) approximation~\cite{PerBurErn-PRL-96} is chosen for $e_c$.
In ref.~\onlinecite{GinSceLooTou-JCP-20} eq.~\eqref{eq:reg} is seen as a short-range functional of the density and of the on-top pair density.

Eq.~\eqref{eq:reg} is the working equation of the PBE-OT method (OT stands for the on-top pair density, $P_2(\bfR,\bfR)$).
Note that eq.~\eqref{eq:reg} aims only to correct the behavior at $\mu=0$, not that at large $\mu$.
}

\updated{In this paper, we do not use any regularization, and compare our
results with those obtained using the PBE-OT ansatz
as well as those obtained by complete basis set two-point extrapolation using the cardinal number characterizing the basis set (eq. (7) of ref.~\onlinecite{HelKloKocNog-97}).}{}

\updated{}{
\subsection{Choosing the basis-set error defining parameter}
\label{subs-def-muB}
}
\updated{}{Thus far, we have shown that as $r \rightarrow 0$, the behavior of the basis-set calculation exhibits a similarity to that of the model characterized by $\mu_\epsilon$, eq.~\eqref{eq:mu-eps}.
However, we have not established a specific relationship between the Gaussian broadening parameter $\epsilon$ and a given basis set.
For the future, we intend to explore the construction of $\mu_\epsilon$ step by step, following the argumentation presented above.}
For the moment, we adopt a simpler strategy: we ``import'' the definition of $\mu_\epsilon$ given in eq.~(37) of ref.~\onlinecite{GinPraFerAssSavTou-JCP-18}.
To make clear that it is just an ``inherited'' definition, we denote it by $\mu_B$, akin to the notation given in that paper.
There are two reasons for using this definition.
First, $\mu_B$ was defined by a projection of the electron-electron interaction on the basis at $r=0$.
This is close to what we saw in eq.~\eqref{eq:w-proj}.
Second, we can study the effect of eliminating the density functional regularization, eq.~\eqref{eq:reg}, and use the ABC expression, eq.~\eqref{eq:bsc} instead.
The regularized expression was used with success in several applications~\cite{LooPraSceTouGin-JPCL-19,GinTraPraTou-JCP-21,TraTouGin-JCP-22,TraTouGin-24}
and we can study if ignoring the regularization and staying simply with the leading order asymptotic expression affects the quality of the results.

In order to facilitate the understanding of our paper, we recall a few elements of the how the $\mu_B$ was constructed in ref.~\onlinecite{GinPraFerAssSavTou-JCP-18}.
\updated{}{ 
First, the Coulomb operator is projected exactly on the basis $B$ of orthonormal molecular orbitals, $\phi$.
Next, it is weighted by the two-particle density, yielding
\begin{equation}
 \label{eq:f}
 f(\bfr_1,\bfr_2) = \sum_{i,j,k,l,m,n} P_{m,n,k,l} \phi_m(\bfr_1) \phi_n(\bfr_2) \phi_i(\bfr_1) \phi_j(\bfr_2) \int_{\R^3} \dd \bfr_3 \int_{\R^3} \dd \bfr_4 \; \phi_i(\bfr_3) \phi_j(\bfr_4)  \frac{1}{|\bfr_3-\bfr_4|} \phi_k(\bfr_3) \phi_l(\bfr_4).
\end{equation}
$P_{m,n,k,l}$ are the coefficients of the two-body density matrix in the basis of the $\phi$.
The indices $i,j$ originate from the projection on the basis for the two particles.
By integration over $\bfr_1$ and $\bfr_2$, $f$ produces the same value as the expectation value of the Coulomb operator in the basis $B$.
However, the Coulomb operator projected on the basis does not possess a singularity at $r=0$.
In order to simplify the treatment, one compares this integrand with the one obtained when using $w(r,\mu)$, eq.~\eqref{eq:w}, namely $w(|\bfr_1-\bfr_2|, \mu) P_{2,B}(\bfr_1,\bfr_2)$ where $P_{2,B}(\bfr_1,\bfr_2)$ is the pair density in the basis.
As one is interested to characterize the short range-behavior, one sets the two expressions equal at $\bfr_1=\bfr_2=\bfR$, obtaining from $w(0,\mu)=(2/\sqrt{\pi}) \mu$ 
\begin{equation}
 \label{eq:def-mub}
 \mu_B(\bfR) = \frac{\sqrt{\pi}}{2} \frac{f(\bfR,\bfR)}{P_{2,B}(\bfR,\bfR)} .
\end{equation}
}


\updated{
To determine $\mu(\bfR)$, one first defines the operator
}{}
\begin{equation}
\label{eq:f-op}
  \updated{\msout{f= \mathcal{P}_B(1,2) w  \mathcal{P}_B(1,2) P_2 \mathcal{P}_B(1,2)}}{\nonumber}
\end{equation}
\updated{
where
$\mathcal{P}_B(1,2) = \mathcal{P}_B(1) \otimes \mathcal{P}_B(2)$
and $\mathcal{P}_B(i)$ is the projector on the orbital basis for
electron $i=1,2$.
For a set of orthonormal orbitals $\phi$, the corresponding two-particle product basis is also orthonormal, and the projector can be written explicitly as:
}{}
\[ \updated{\msout{
P_B(1,2) = \sum_{i \in B} \sum_{j \in B} \phi_i(\mathbf{r}_1)\phi_j(\mathbf{r}_2) \phi_i(\mathbf{r}'_1)\phi_j(\mathbf{r}'_2).
}}{} \]
\updated{
Hence,
}{}
\[ \updated{\msout{
  f(\bfr_1,\bfr_2,\bfr_3,\bfr_4) = \sum_{i,j,k,l,m,n} \phi_i(\bfr_1) \phi_j (\bfr_2) \langle \phi_i \phi_j | w | \phi_k \phi_l \rangle \langle \phi_k \phi_l | {P}_2 | \phi_m \phi_n \rangle \phi_m(\bfr_3) \phi_n(\bfr_4).
}}{} \]
\updated{
If one considers that $\mathcal{P}_B(1,2) w  \mathcal{P}_B(1,2)$ can
be replaced by a model $w$, depending on some $\mu$, cf. eq.~\eqref{eq:w-proj}, for $\bfr_1=\bfr_2=\bfr_3=\bfr_4=\bfR$,
we get for each $\bfR \in \R^3$ a model parameter $\mu_B$,
}{
}

\updated{}{\subsection{Working equation for this paper}
\label{subs:working-eq}

In this work, we equate  our $\mu_\epsilon$ with $\mu_B$ defined as above.
So,  our working equation in this paper is obtained by replacing $\mu_\epsilon$ by $\mu_B$ in eq.~\eqref{eq:bsc},}
\begin{equation}
 E \approx \expect{H}{\Psi_B} -   \frac{4 (\sqrt{2} -1)\sqrt{\pi}}{6}\int_{\R^3} \dd \bfR \, P_{2,B}(\bfR,\bfR) \left( 1 + \frac{2}{\sqrt{\pi} \mu_B(\bfR)} \right)^{-1}    \mu_B(\bfR)^{-3} ,
 \label{eq:e-abc} 
\end{equation}
\updated{}{
where $\Psi_B$ is the FCI wave function obtained with the basis set $B$, and $P_{2,B}$ is the pair density generated by it.

In practice,
\begin{itemize}
 \item we perform a selected CI calculation in a given basis to obtain the expectation value of the physical Hamiltonian with the wave function $\Psi_B$ obtained in this basis,
 \item we compute the on-top pair density on a three-dimensional grid,
 \item we compute $\mu_B$ on the same grid points,
 \item we numerically evaluate the integral in eq.~\eqref{eq:e-abc}.
\end{itemize}

\updated{}{As, in this paper, we correct calculations in Gaussian basis sets, $\mu_B$ is considered to depend on the electron-electron coalescence position $\bfR$.
Using a global basis set, such as one consisting on plane waves, eliminates this dependence of $\mu_B$ on $\bfR$.
A position-dependence remains: the basis-set correction requires only the spatial average of the on-top pair density.
The latter can be also estimated from expectation values of short-range two-particle operators~\cite{{SceSav-Trygve-JPCA-24}}.
}

At this point, we would like to make a comparison with range-separated density functional methods.
Using a local parametrization of a range-separation parameter has been shown to improve the results (see, \textit{e.g.}, refs.~\onlinecite{PolSavLeiSto-JCP-02, KruScuPerSav-08}).
In the present case, it is the basis set used that defines the local range-separation parameter.
Thus, one avoids calculating the expectation value of a model Hamiltonian with a two-particle operator that does not depend only on the distance between particles.
Eq.~\eqref{eq:e-abc} only requires a three-dimensional numerical integration, not computing a new type of two-electron integrals.
}

\section{Numerical results}

\subsection{Calculation details}

In the following, we first \updated{strengthen}{support} some of the statements made above using some numerical results.
Next we show how the method works for some selected systems.
In all these examples we will present results from
\begin{itemize}
 \item[a)] full configuration interaction (FCI) as estimated from
   selected configuration interaction calculations,
 \item[b)] the two point extrapolation formula to the complete basis set (CBS),
 \item[c)] the method using the on-top pair density and the PBE density functional (PBE-OT)~\cite{GinSceLooTou-JCP-20}
 \item[d)] ABC, eq.~\eqref{eq:e-abc}.
\end{itemize}
The accuracy of FCI energies estimated from selected CI is \updated{beyond}{better than} chemical accuracy.

In this paper, we adopt the following CBS expression:
\begin{equation}
 E_X \approx a + b \, X^{-3}
\label{eq:cbs}
\end{equation}
where $X$ is the cardinal number ($X$ in a V$X$Z type basis set~\cite{Dun-JCP-89}).
$E_X$ is the energy obtained from a calculation using the V$X$Z basis.
$a$, and $b$ are obtained from a fit to two calculations, $X$, and $X-1$.
$a$ is the estimate of the energy in the complete basis set limit.
Our CBS extrapolation formula differs from the original expression, eq. (7) of ref.~\onlinecite{HelKloKocNog-97}, that applies this ansatz only to the correlation, not to the total energy, $E$.
When the basis set is large enough to ensure proximity of the Hartree-Fock calculation to the Hartree-Fock limit, the difference is just a shift of the constant $a$.
CBS cannot be used to correct the energy \updated{only from the smallest basis set}{from the smallest basis set alone}, as two calculations are needed for the extrapolation.
In this case, it appears \updated{in the plots identical to the FCI result}{identical to the FCI result in the plots}.

Our results include those obtained for Harmonium, that is for a one-particle potential
\begin{equation}
 \label{eq:v-harm}
 v(\bfR) = \frac{1}{2} \omega^2 R^2
\end{equation}
where we chose $\omega=1/2$, and considered states for $N=2$ or $N=4$ electrons.
We are not aware of V$X$Z basis sets for these systems, so we constructed even-tempered basis sets as analogs to VDZ ($X=2$) basis sets: a $3s1p$, for $N=2$, and a $5s3p1d$, for $N=4$.
In the spirit of ref.~\onlinecite{Dun-JCP-89} we progressively added, for each angular momentum present, another basis function and a supplementary basis function with the next higher angular momentum.
The basis-set parameters can be found in the Supplementary Information.

For the ground state of $N=2, \omega=1/2$ analytical results are known (see, \textit{e.g.}, ref.~\onlinecite{Kin-TCA-96}).
For example,
\[ \Psi  = \frac{1}{2 \pi ^{3/4} \sqrt{\left(8+5 \sqrt{\pi }\right) \pi } } \left( 1 + \frac{r}{2} \right)  e^{-\frac{r^2}{8}} e^{-\frac{R^2}{2}} \]
and
\begin{equation}
 P_2(\bfR,\bfR) = \frac{1}{16 \pi^{5/2} +10 \pi ^3} e^{-R^2}
\label{eq:ontop-harm}
\end{equation}

CI calculations were made with the \textsc{Quantum Package} program~\cite{GarGasBen-JCTC-19}.
Symbolic calculations were performed by using Mathematica~\cite{Mathematica}.
\updated{Reference energies for the studied systems are given in table~\ref{tab:ref-energies}.
Differences between our computed and the reference energies can be found in the Supplementary Information.}{
Reference energies of the studied systems and our computed energies can be found in
the Supplementary Information.
These also contain tables giving $\mu$ as a function of $R$ for Harmonium and the He atom.}

\subsection{Short-range behavior of the wave function}

\begin{figure}[ht]
 \includegraphics{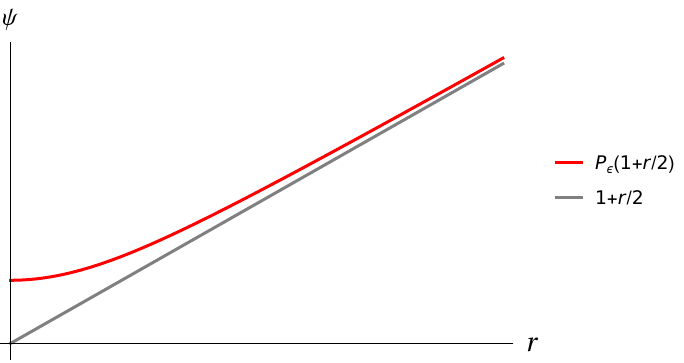}
 \caption{Schematic behavior of the wave function at short distances between electrons: exact, satisfying Kato's cusp condition eq.~\eqref{eq:psi-l}, gray, and projected on a finite basis set with an approximate projector, eq.~\eqref{eq:proj-kato}, red, both for $\ell=0$.
}
\label{fig:psi}
\end{figure}
The short-range behavior of $\psi_0(r)$, eq.~\eqref{eq:psi-kato}, is illustrated schematically in fig.~\ref{fig:psi}.
At the origin, the physical $\psi_0$ starts linearly with $r$.
The projection with $\mathcal{P}_{2 \epsilon}$, eq.~\eqref{eq:proj-kato}, modifies $\psi_0$ to start quadratically ($r \ll 2 \epsilon)$.
For $r \gg 2 \epsilon$, the effect of the projection vanishes, $\mathcal{P}_{2 \epsilon}(\bfr,\bfr') = \delta(\bfr-\bfr')+ \mathcal{O}(\epsilon)$.
$\psi_0(r,\lambda=0,\mu=1/\sqrt{4 \epsilon})$, eq.~\eqref{eq:psi-0-lambda-mu} produces the same curve as the one obtained by projection.

\subsection{Basis-set errors in model calculations}

\begin{figure}[ht]
 \includegraphics{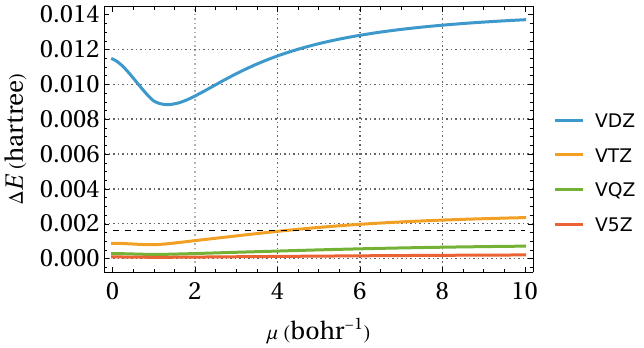}
 \caption{Basis-set errors generated by using (uncontracted) V$X$Z
   basis sets for the He atom.
The horizontal dashed line indicates chemical accuracy (1~kcal/mol).}
\label{fig:bse-of-mu}
\end{figure}
To use eq.~\eqref{eq:e-approx} one must first solve (to satisfactory accuracy) the Schr\"odinger equation for the model Hamiltonian.
The advantage of using the model Hamiltonian (containing a potential without a singularity instead of using the physical Hamiltonian) is a better convergence with the basis-set size.
Fig.~\ref{fig:bse-of-mu} shows estimates of $\expect{H}{\Psi(\lambda=0,\mu)}$ when different basis sets are used.
It shows that, indeed, the basis-set errors are smaller for the model interactions than for the Coulomb interaction.
Furthermore, we see that small values of $\mu$ should be preferred.
However, achieving the desired accuracy may require values of $\mu$ smaller than the range where
eq.~\eqref{eq:e-approx} is valid, which is limited to large $\mu$.

We also notice that for smaller basis sets errors are present even for the non-interacting system (at $\mu=0$).
These are deficiencies of the basis set that we do not intend to correct in the present paper.

In the example shown in fig.~\ref{fig:bse-of-mu}, about half of the error at $\mu=0$ originates from our decision not to change the one-particle potential when modifying the interaction.
We believe that the critical value of $\mu$ below which
eq.~\eqref{eq:e-approx} becomes invalid corresponds to the point where the universality of the short-range behavior breaks down.~\cite{SceSav-JCC-24}
As the present treatment is based on a universal form of the wave
function at small inter-electronic distances, whether or not a
mean-field potential is adopted is considered irrelevant.

\subsection{Validity of the adiabatic correction approach}
\begin{figure}
\includegraphics{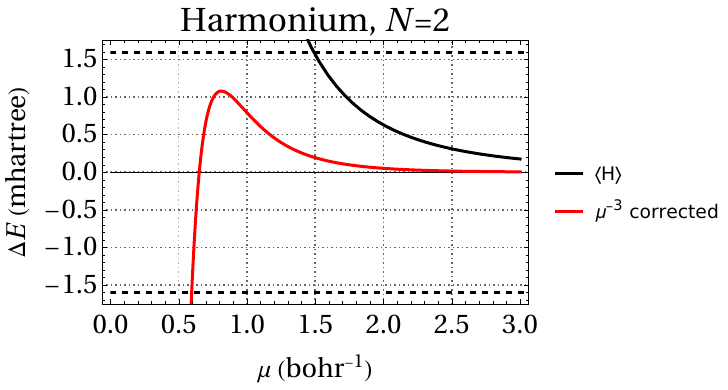} \\
\includegraphics{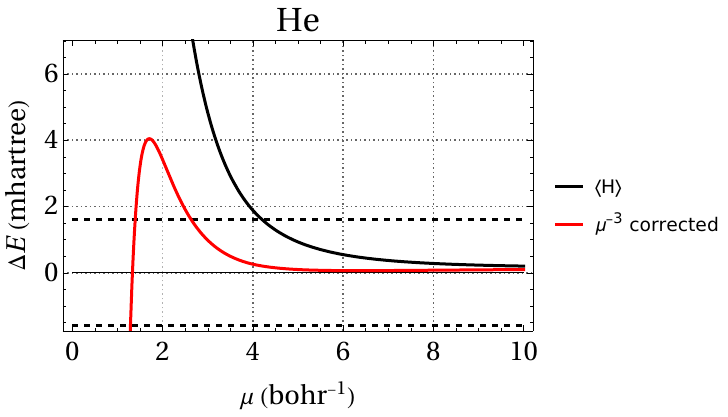}
 \caption{Energy errors (in mhartree), for the ground states of Harmonium, $N=2, \omega=1/2$ (top panel) and of He (bottom panel). The curve showing the errors of $\expect{H}{\Psi(\lambda=0,\mu)}$ is in black, that after correcting with eq.~\eqref{eq:e-approx} is in red. The horizontal dashed lines indicate errors of $\pm 1$~kcal/mol.}
\label{fig:ac-error}
\end{figure}
Fig.~\ref{fig:ac-error} shows errors of
$\expect{H}{\Psi(\lambda=0,\mu)}$, which are always positive.
For larger values of $\mu$ the effect of the correction after using the asymptotic exact expression, up to order $\mu^{-3}$ eq.~\eqref{eq:e-approx} is evident.

We notice in fig.~\ref{fig:ac-error} different scales for Harmonium
and for He, which results from our choice of the model potential.
Given that $\mu$ has the dimension of an inverse length,
it has to adapt to the scale of the region of coalescence: the nuclear charge in He, $Z=2$, makes the system more compact than Harmonium for the chosen parameter ($\omega=1/2$).
The locality of $\mu_B$ should take care of it.

We attribute the appearance of the bump in the corrected curves, fig.~\ref{fig:ac-error}, to two effects.
One is the use of the bare nuclear potential, and not some mean-field potential at $\mu=0$.
This lowers the energy very strongly, but \updated{it}{this}
is not surprising.
Using a regularization such as in eq.~\eqref{eq:reg} helps correcting this effect.
We think that the other effect generating errors is stopping the expansion at the leading term in $1/\mu$; in the treatment exposed in this paper, higher order terms are totally missing.
\updated{In principle, this can also be arranged by regularization.
However, a form as given in}{Regularization as in} eq.~\eqref{eq:reg} \updated{is not sufficiently flexible}{does not solve the problem}.
\[
  \updated{\msout{\frac{a}{1+ b \mu^3} = a -\frac{a \, b}{\mu^3} +\frac{a \, b^2}{\mu^6} + \dots}}{} \]
\updated{because the higher order terms are determined by the simple form chosen for the regularization.}{}
The second type of error is progressive, in contrast to the first one that is drastic.

The critical value,  $\mu_{\text{crit}}$, below which our correction eq.~\eqref{eq:e-approx} fails varies from system to system, so we must use a basis set for which $\mu_B(\bfR) > \mu_{\text{crit}}$.
Let us require that the absolute error produced by the adiabatic connection stays below 1~kcal/mol.
For example, for the Harmonium, $N=2, \omega=1/2$ ground state (fig.~\ref{fig:ac-error}, top panel) we have $\mu_{\text{crit}} \approx 0.6$~bohr$^{-1}$.
For He (fig.~\ref{fig:ac-error}, bottom panel), $\mu_{\text{crit}} \approx 2.7$~bohr$^{-1}$.
Note, however, that the error we attribute the missing higher order terms to (before the failure becomes catastrophic) remains relatively small for $\mu > 1.3$~bohr$^{-1}$.

\begin{figure}
\includegraphics[]{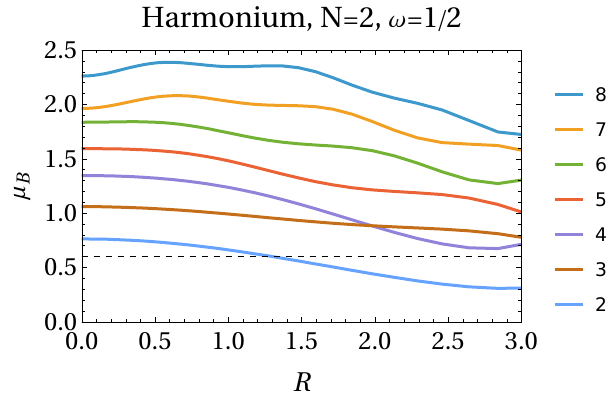} \\
\includegraphics[]{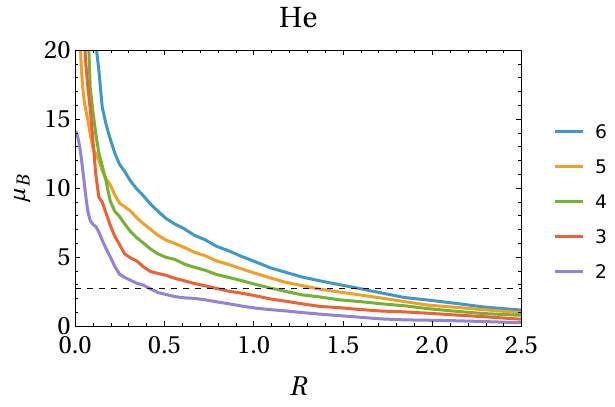}
\caption{The model parameter, $\mu_B$, for different V$X$Z-type basis sets, as a function of $R$; $X=2$ to $X=8$ (Harmonium $N=2, \omega=1/2$, top panel; He bottom panel). The horizontal dashed line shows the critical value of the parameter, $\mu_{\text{crit}}$ below which the asymptotic expansion fails (see fig.~\ref{fig:ac-error}).}
\label{fig:mu-of-R}
\end{figure}

We present in fig.~\ref{fig:mu-of-R}, top panel, $\mu_B(R)$ for Harmonium.
It exhibits a weak dependence on the position of the coalescence point for the Harmonium ground state.
For Harmonium, all basis sets V$X$Z, $X \ge 3$ produce $\mu_B(R) > \mu_\text{crit}$.
Using the VDZ basis set, $\mu_B(R)$ gets below $\mu_\text{crit}$ for large $R$.
Nevertheless, $X=2$ seems to be at the limit of applicability, because $P_2(\bfR,\bfR)$, eq.~\eqref{eq:ontop-harm}, decreases fast at large $R$, and we do not expect this violation to have an important effect.

We expect $\mu_B$ to increase with the basis-set cardinal number, $X$.
This is true, except for large $R$ (small density, lower weight in obtaining the energy) \updated{}{where we find an inverted order} between $X=3$ and $X=4$.
This can be rationalized by the construction of our basis sets.
We used even-tempered basis sets, and the larger basis set includes the smaller basis set only for $X$ of the same parity.

The behavior of $\mu_B(R)$ for He is different (bottom panel of fig~\ref{fig:mu-of-R}).
First, we see that it decays fast with $R$.
We suspect that this is related to the energy optimization of the V$X$Z basis sets that puts emphasis on the regions close to the nucleus.
Even larger basis sets seem not to deliver $\mu_B(R) > \mu_\text{crit} $ for large $R$.
Again, the importance of what appears to be a deficiency of the basis set, is not so important as it occurs for large $R$, where $P_2(R,R)$ is small.
Furthermore, if we consider the region where $\mu$ is still larger than the one where the drastic change in the error produced in our approximation shows up ($\approx \mu_\text{crit}/2$), the critical region is pushed to higher $R$, and thus has less weight.
We also observe some crossings, and we believe that this could be a reflection of the fact that the basis set with larger $X$ does not contain the basis functions of the basis set with smaller $X$.
Nevertheless, the general behavior is that $\mu(R)$ increases with the basis-set size.

We do not pursue this analysis for other systems.
Nevertheless, it shows up in the energy errors obtained with different basis sets for different systems presented further down.

\subsection{Effect of the basis set on the on-top pair density and its correction}

\begin{figure}
\includegraphics[]{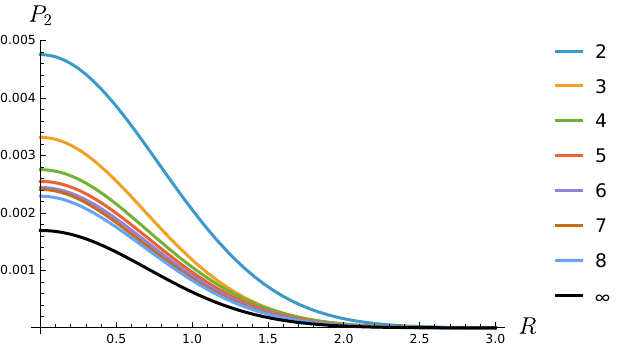} \\
\includegraphics[]{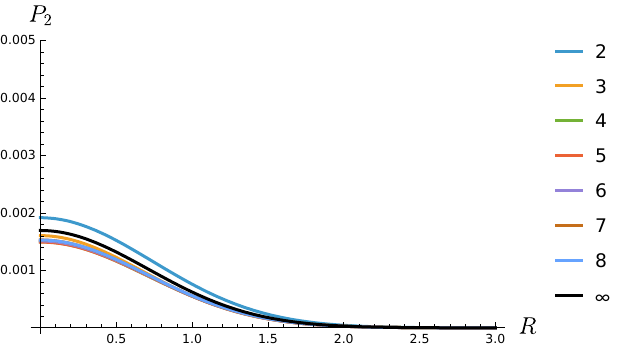}
\caption{On-top pair density with different basis sets before (top
  panel) and after (bottom panel) correction using eq.~\eqref{eq:p2-mu}, for Harmonium, $N=2, \omega=1/2$ ground state. The black curve is the exact one, while the colored curves correspond to our V$X$Z basis sets ($X$ indicated as legend).}
 \label{fig:ontop}
\end{figure}

It is well known that the on-top pair density is very sensitive to the basis set used, and that its value converges slowly to the exact value (see, \textit{e.g.}, Table IV of ref.~\onlinecite{Dav-63} for its system-average for He, given as $ \langle \delta(\bf r_{12}) \rangle $).
This is confirmed for the ground state of Harmonium ($N=2,
\omega=1/2$~a.u. (fig.~\ref{fig:ontop}, top panel).
To facilitate the comparison, we utilize the known exact form of $P_2(\bfR,\bfR)$, eq.~\eqref{eq:ontop-harm}.
We see that even with the largest of the basis sets the error is still large.

After correction, eq.~\eqref{eq:p2-mu}, we see that the on-top pair
density is much improved (fig.~\ref{fig:ontop}, bottom panel).
With the exception of the smallest of the basis sets (VDZ), the difference between the exact, eq.~\eqref{eq:ontop-harm},  and corrected, eq.~\eqref{eq:p2-mu}, on-top pair density  is hard to distinguish on the scale of our plot.

\subsection{Connection between CBS and $\mu^{-3}$ dependence}

\begin{figure}
 \includegraphics[]{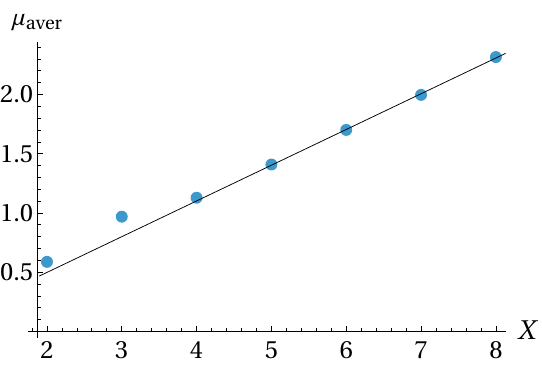}
 \caption{Correlation between the cardinal number characterizing the V$X$Z basis, $X$ and an system-averaged value of $\mu$ for the Harmonium $N=2, \omega=1/2$ ground state.}
 \label{fig:n-mu-correlation}
\end{figure}

The estimated energy in CBS limit ($a$ in eq.~\eqref{eq:cbs}) should
be correct to order $X^{-3}$.
(We can even expect some higher order contributions through the fitting to a pair of data.)
This raises the question of whether a correlation exists between the parameter $\mu$ characterizing the system and the basis-set cardinal number, $X$.
To see it, we consider again the Harmonium $N=2,\omega=1/2$ ground state, for which $P_2(\bfR,\bfR)$ is given in eq.~\eqref{eq:ontop-harm}.
We define a system-averaged value of $\mu$ as $\mu_{aver}$ obtained using eq.~\eqref{eq:mu-correction}, and the exact form of $P_2(\bfR,\bfR)$, eq.~\eqref{eq:ontop-harm},
\[
\int_{\R^3} \dd \bfR \, P_2(\bfR,\bfR)   \mu_B(\bfR)^{-3}
= \mu_{aver}^{-3}  \int_{\R^3} \dd \bfR \, P_2(\bfR,\bfR)
\]
Fig.~\ref{fig:n-mu-correlation} shows indeed a good linear correlation, especially for large $\mu$.
This correlation might point \updated{out}{} to a connection between  the construction of the V$X$Z basis sets and the CBS extrapolation, that is, the $X^{-3}$ behavior, eq.~\eqref{eq:cbs}, and a model (such as ours, using $\mu$) in the complete basis set limit for the model calculation.

\subsection{Energy errors for some selected systems}

\begin{figure}
\includegraphics{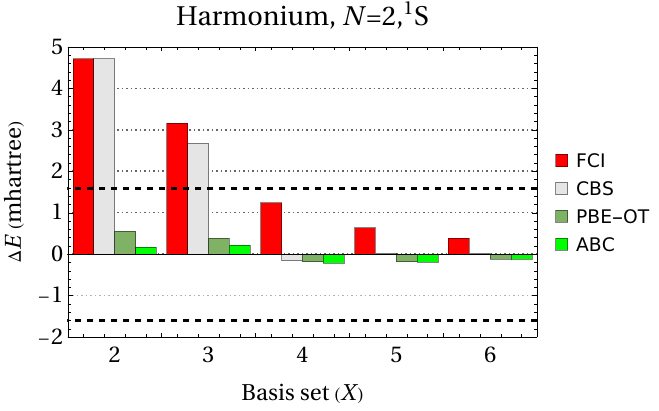} \\
\includegraphics{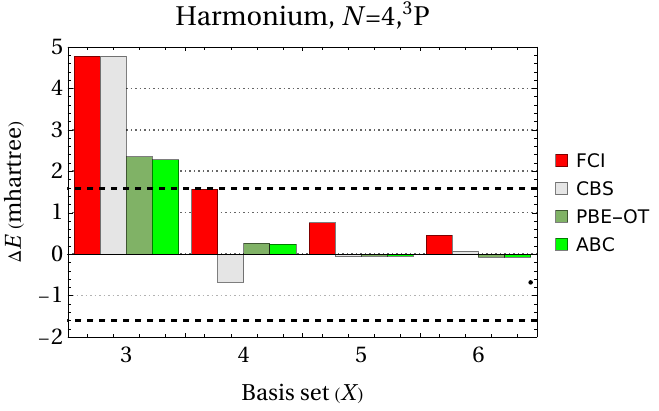}
 \caption{Energy errors (in mhartree), for Harmonium ground states, $\omega=1/2$, $N=2$ (top) and $N=4$ (bottom), for basis sets of increasing size; from left to right, FCI, red bars, after CBS extrapolation, gray bars, after PBE-OT correction, dark green bars, and using eq.~\eqref{eq:e-abc}, light green bars. The horizontal dashed lines indicate chemical accuracy, $\pm 1$~kcal/mol.}
\label{fig:de-harm-gs}
\end{figure}
We now analyze the correction of the basis-set errors for specific systems.
The results for Harmonium, $N=2,^1$S and $N=4$, $^3$P ground states are shown in fig.~\ref{fig:de-harm-gs}.
In all cases the correction given either by the PBE-OT approximation or eq.~\eqref{eq:e-abc} gives an improvement over the FCI result.
The ABC results are comparable.
Except for the smallest of the basis sets, the errors are below
1~kcal/mol, the so-called chemical accuracy~\cite{Pop-RMP-99}.
For $X \ge 4$ they are also comparable in magnitude to the CBS result.
(The CBS results seem slightly superior for larger $X$, maybe due to fitting to a pair of data.)
Errors associated with small basis sets warrant a separate discussion below.

\begin{figure}
\includegraphics[width=0.45\textwidth]{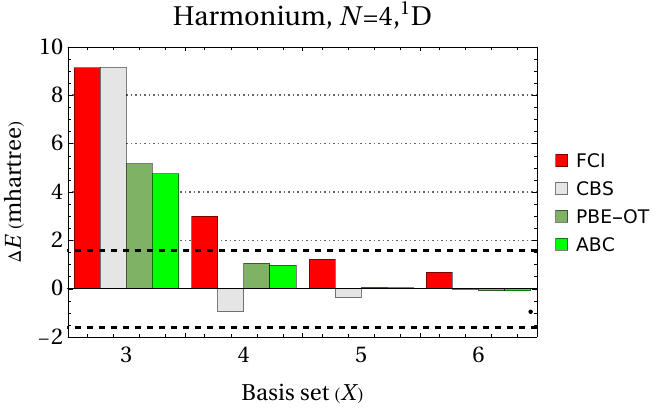}
\includegraphics[width=0.45\textwidth]{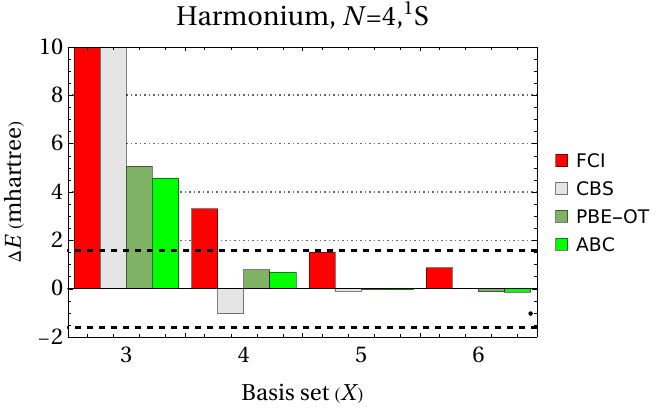} \\
\includegraphics[width=0.45\textwidth]{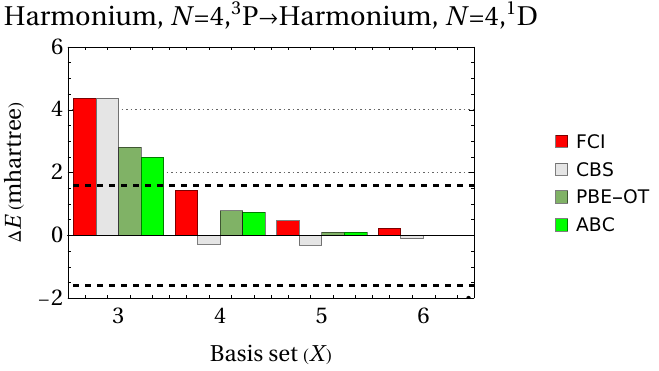}
\includegraphics[width=0.45\textwidth]{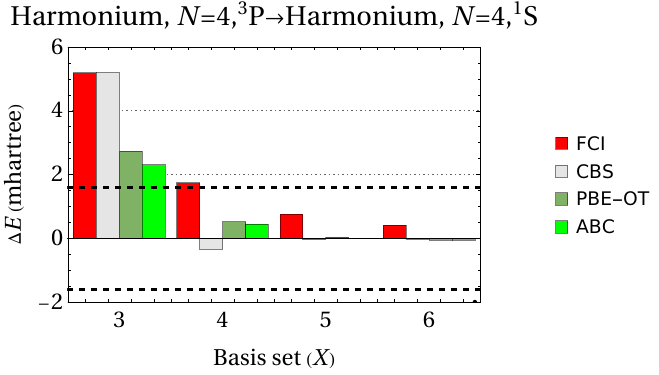}
 \caption{Energy errors (in mhartree), for Harmonium excited states, $\omega=1/2$, $N=4$ for the $^1$D state (left) and $^1$S state (right), and basis sets of increasing size. The errors of the total energies are on the top, those of the excitation energies on the bottom. Further details as in fig.~\ref{fig:de-harm-gs}}
\label{fig:dde-harm-excit}
\end{figure}
Fig.~\ref{fig:dde-harm-excit} shows the errors in excited states, and excitation energies.
Some error compensation occurs for the excitation energies, already at
the FCI level.
Except for the smallest basis sets, the errors are below 1~kcal/mol for all the corrected data.

\begin{figure}
\includegraphics{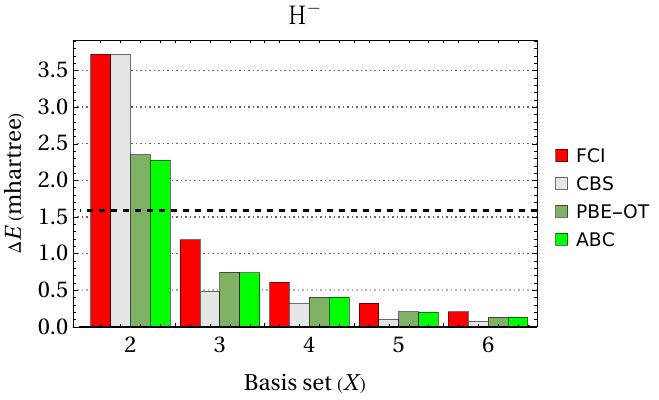} \\
\includegraphics{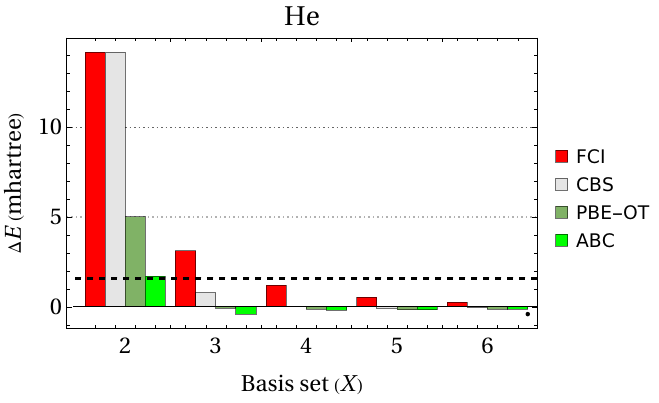} \\
\includegraphics{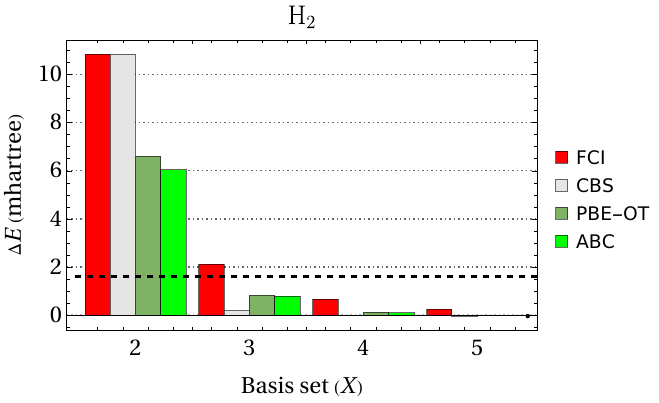}
  \caption{Energy errors (in mhartree), for two-electron systems, in their ground states, H$^-$(top), He (center), and H$_2$, \updated{$^1\Sigma_u$}{$^1\Sigma_g$} state, at equilibrium distance (bottom), for V$X$Z basis sets. Further details as in fig.~\ref{fig:de-harm-gs}. }
 \label{fig:2el-atmol-gs}
\end{figure}
Fig.~\ref{fig:2el-atmol-gs} shows the results for the ground state of some two-electron systems, H$^{-}$, He, and H$_2$ at equilibrium distance.
We observe large FCI errors for He and H$_2$ with the VDZ basis set.
For $X>2$ chemical accuracy is obtained for all systems after correction (for H$^{-}$ already at FCI level).

\begin{figure}
\includegraphics{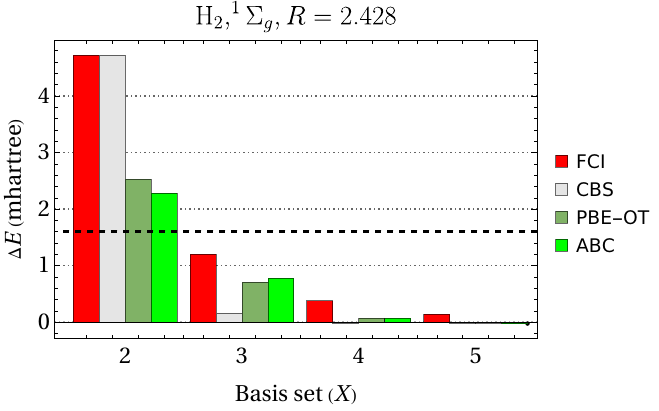} \\
\includegraphics{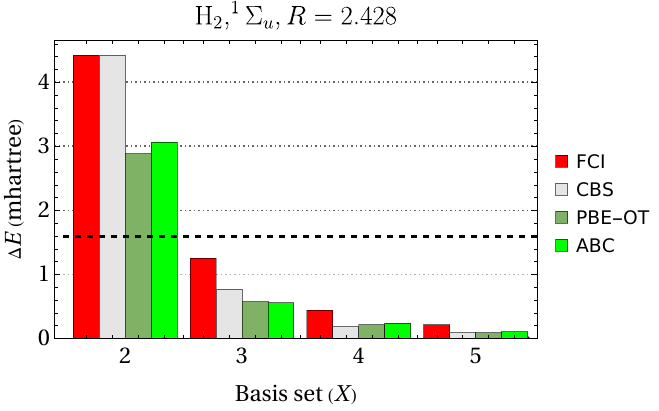} \\
\includegraphics{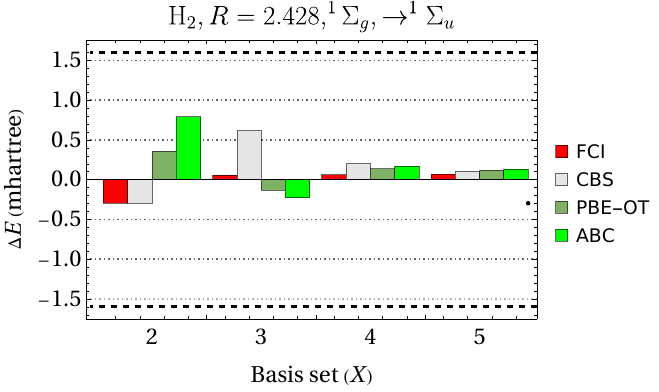} \\
 \caption{Energy errors (in mhartree), for H$_2$ at the internuclear
   separation corresponding to the minimum of the $^1\Sigma_u$ state
   ($R=2.428$~bohr), for the $^1\Sigma_g$ state (top), the $^1\Sigma_u$ state (center) and the energy difference between them (bottom), for basis sets of increasing size. Further details as in fig.~\ref{fig:de-harm-gs}.}
 \label{fig:2el-atmol-excit}
\end{figure}
We also consider the first excited singlet state of the H$_2$ molecule, a $^1\Sigma_u$ state (fig.~\ref{fig:2el-atmol-excit}).
Total energies do not attain chemical accuracy for the VDZ basis set, whereas other basis sets do, with or
without correction.
For the vertical excitation energy to the minimum of the $^1\Sigma_u$ state, chemical accuracy is reached also with the VDZ basis set, for FCI, with or without correction.

\begin{figure}
\includegraphics{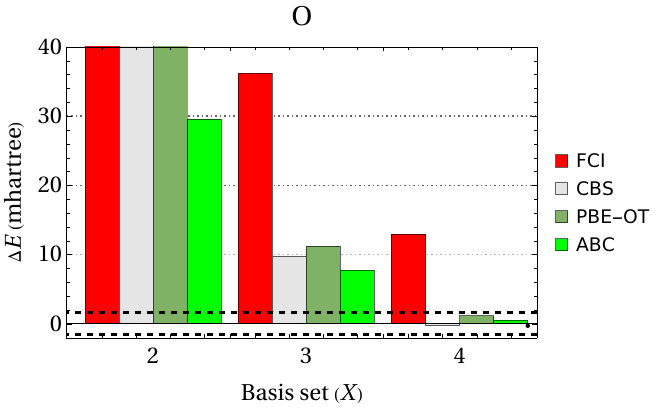} \\
\includegraphics{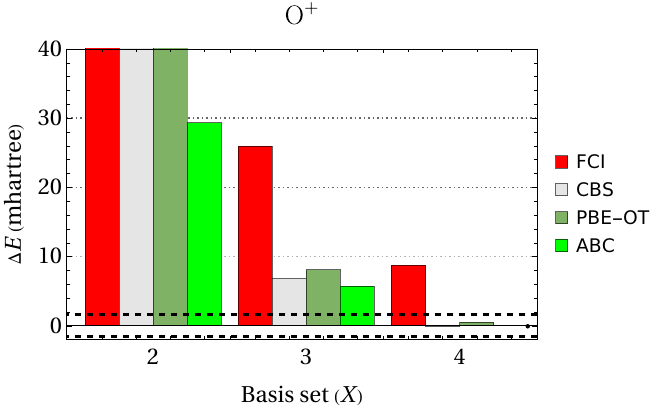} \\
\includegraphics{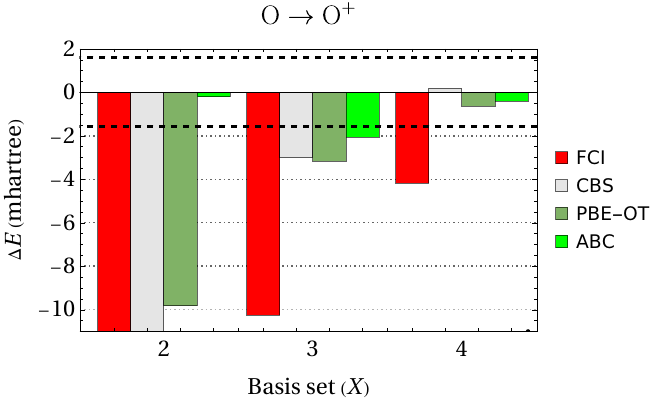}
 \caption{Energy errors (in mhartree), for the O atom (top), for its
   ion, O$^+$, and for the ionization potential (bottom) for V$X$Z
   basis sets.
   Further details as in fig.~\ref{fig:de-harm-gs}.}
 \label{fig:o}
\end{figure}
Fig.~\ref{fig:o} shows the worst case from our test examples.
For oxygen, at VDZ levels the total energy errors are huge.
They do not reach the desired accuracy even for the VTZ basis set.
This is also the case for the O$^+$ ion.
Some error compensation is present in the ionization potential, \updated{but the error remains larger than the one set as our objective of chemical accuracy}{but the error remains larger than our target of chemical accuracy}.
\updated{Only for VQZ we can reach chemical accuracy, and this after basis-set error correction}{Chemical accuracy is reached only for VQZ, and only after basis-set error correction}.
Nevertheless, we observe a significant improvement of the ionization energy after applying the correction.

\subsection{Hartree-Fock and correlation energy errors}
\label{subs:HF-errs}

\begin{figure}
\includegraphics{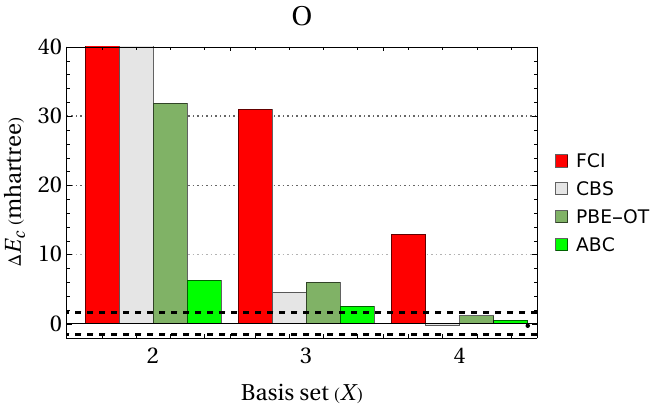} \\
\includegraphics{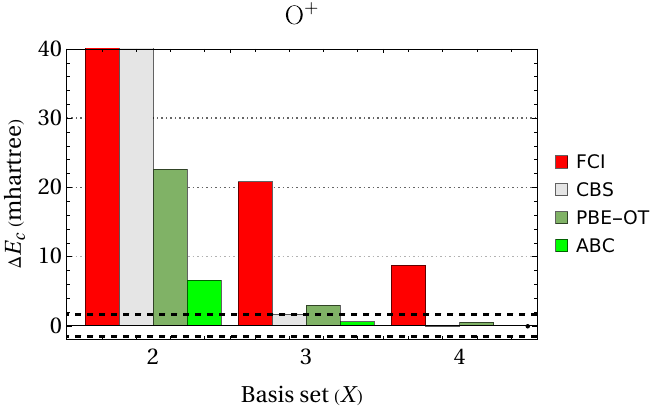}
 \caption{Correlation energy errors (in mhartree), for the O atom
   (top), and for its ion, O$^+$ (bottom), for V$X$Z basis sets.
   Further details as in fig.~\ref{fig:de-harm-gs}.}
 \label{fig:dec-o}
\end{figure}

As already mentioned in the introduction, the correction presented in this paper is specifically designed to address errors in the correlation energy arising from the poor description of the wave function when $r \rightarrow 0$ and other types of basis-set errors may exist.
In particular, those at Hartree-Fock level are not treated by ABC.
These errors can be significant when using smaller basis sets.
For example, the difference between our Hartree-Fock energy of the oxygen atom with the VDZ basis set and the V6Z \updated{to approximate the Hartree-Fock limit is of}{result, used to approximate the Hartree-Fock limit, is} $\approx 25$~mhartree.
\updated{The VTZ basis set}{With the VTZ basis set,} this improves significantly, but an error of $\approx 7$ mhartree persists.
This limitation rationalizes the larger total energy errors observed in our results when using smaller basis sets, even after applying the ABC correction.

To isolate the performance of our correlation correction, we examine the correlation energy errors alone (Fig. \ref{fig:dec-o}). Comparing these results with the total energy errors in Fig. \ref{fig:o} reveals an improvement.
The discrepancy highlights that a high-quality description of the one-electron problem (the HF limit) is a prerequisite for the proposed correction to yield optimal results for total energies.

We did not focus exclusively on correlation energy errors in the main results for two reasons. First, achieving a good description of the one-electron problem is generally straightforward compared to treating correlation; thus, the bottleneck for accuracy usually lies in the correlation treatment, which our method targets. Second, the definition of correlation energy itself can be ambiguous. Depending on whether one adopts a restricted~\cite{Low-58} or unrestricted~\cite{PopBin-MP-75} Hartree-Fock reference, the resulting correlation energy may differ.
Furthermore, spatial symmetry breaking is possible, and one can prefer to avoid it (as, for example, in ref.~\onlinecite{ChaGwaDavParFro-PRA-93}).
Along these lines, a proper description of a configuration may \updated{require} more than a single Slater determinant (as, \textit{e.g.}, for the $^1\Sigma_u$ state of the H$_2$ molecule treated in this paper).

An alternative strategy to mitigate Hartree-Fock errors is to consider energy differences, such as ionization potentials or excitation energies and rely on error compensation.
For example, the Hartree-Fock energy errors for the oxygen atom and its cation (O$^+$) are comparable, leading to a partial cancellation when computing the ionization potential, see fig.~\ref{fig:o}.

\updated{}{
\section{Final remarks}
\subsection{Summary}
}
In this paper, we propose a method to correct the energy errors due to the absence of the electron-electron cusp in the wave function.
We observe that \updated{using finite basis sets for orbitals provide wave functions}{using finite orbital basis sets provides wave functions} that, at short range,  are close to those obtained with a complete basis set for a system with a model interaction (without singularity at the coalescence point of a pair of electrons).
Furthermore, \updated{we obtain from the on-top pair density of the model system that of the physical system}{we obtain the physical-system on-top pair density from the model-system on-top pair density}.
\updated{The adiabatic connection letting the model system evolve to the physical system}{The adiabatic connection that evolves the model system into the physical system} provides an expression for the energy correction.
We call it ``Adiabatic connection based Basis-set error Correction'' (ABC).
This correction is valid when the model approaches the physical (Coulomb) interaction.

This is a general approach.
In order to have a simple practical procedure, we characterize our model  by a number, $\mu$.
This parameter is \updated{a}{} related to the point where the singularity in the interaction is (smoothly) cut off.
It has the dimension of an inverse distance, and approaches the physical system as $\mu \rightarrow \infty$.
With ``local'' basis sets, like Gaussian type orbitals, it varies in space.
\updated{In order to compare with results from the literature, the PBE-OT method, we choose the same spatial dependence of $\mu$}{To compare with the PBE-OT method from the literature, we choose the same spatial dependence of $\mu$}, eq.~\eqref{eq:def-mub}.
For the adiabatic connection, we keep only the leading term of the expansion in $1/\mu$.
\updated{It provides a simple form for the energy correction, eq. (14) and the pair density, eq. (15). and this gives us a simple expression for the correction of the energy error due to the finite basis set.}{This provides simple forms for the energy correction, eq. (14), and the pair density, eq. (15), and gives us a simple expression for the correction of the energy error due to the finite basis set.}
\updated{This provides us a correction of the model}{This provides a correction to the model} in the form of an analytical formula, depending on our local $\mu$ and the pair density generated in the calculation with a finite basis set, eq.~\eqref{eq:e-abc}.
\updated{}{ABC (unlike F12 methods) does not produce wave functions.
Instead it focusses on expectation values.}

The preliminary numerical results are encouraging:  we can reach chemical accuracy when the basis set ensures good Hartree-Fock results.
The results obtained are comparable with those obtained from extrapolation to the complete basis set limit (CBS), or with a correction using the density and the pair density (PBE-OT).
The advantage over CBS is that only a single calculation is needed.
\updated{CBS has the advantage that fitting to two calculations}{CBS can be advantageous because fitting to two calculations} with different basis sets can produce slightly better results than ours.
Numerically, the results from PBE-OT and ABC are similar.
\updated{}{The advantage over PBE-OT is more conceptual: ABC avoids the choice of a density functional and is formally more naturally extendable to excited states and properties.}
By relying solely on a Taylor expansion, we avoid the dilemma of selecting the 'best' density functional.

\updated{}{
  \subsection{Perspectives}
}
Several aspects of ABC require further elucidation or extension.
\begin{enumerate}
  \item The ansatz for the projection operator, $\mathcal{P}_\epsilon$, eq.~\eqref{eq:proj-eps}, has the advantage \updated{to be}{of being} simple, but is far from perfect.
 For example, it is not idempotent.
 Nevertheless, this can be improved.
 \item The connection between $\mathcal{P}_\epsilon$ and the projector on the basis set, $\mathcal{P}_B$, ref.~\onlinecite{GinPraFerAssSavTou-JCP-18} has some degree of arbitrariness.
 This could be exploited either by performing accurate calculations, or proposing different, computationally advantageous forms.
 \item We stopped at the leading term in the expansion for large $\mu$.
 Going further is possible, and some improvement can be expected (see, \textit{e.g.}, fig. 2 in ref.~\onlinecite{KarSav-MP-22}).
 This might improve results for smaller basis sets, \textit{e.g.}, the He atom for the VDZ basis set.
 \item Because it appears in higher orders, same-spin correlation is not taken into account by our correction, but is
 supposed to be described by the expectation value of the Hamiltonian.
 Of course, this is not exact.
 We have estimates of this effect in fig.~13 of ref.~\onlinecite{SceSav-Trygve-JPCA-24} and it did not appear important to us for the systems studied so far.
 Furthermore, \updated{we might have to treat first terms in $\mu^{-4}$ before we treat the terms in $\mu^{-5}$}{we may have to treat the leading $\mu^{-4}$ terms before treating the $\mu^{-5}$ terms}, that show up with same-spin contributions.
 \updated{}{Nevertheless, this point has to be studied more carefully.}
 \item The short-range behavior is universal.
 This is why, up to now, we did not take into account mean-field potentials.
 It seems that in this regime they are not needed.~\cite{SceSav-JCC-24}
 However, this might become an issue for larger systems.
 \item
   \updated{}{
 Our derivation is valid only for large values of $\mu$.
 However, we do not know when this regime is valid.
 This kind of problem is common to practically all problems in quantum chemistry: we do not have error bounds to our results.
 This applies even in situations where there are proofs that the exact results are approached systematically (as for variational method, the extrapolation using partial wave expansion, that is, using complete basis sets for angular momenta up to $L$, leading to the  $L^{-3} + \dots$ behavior).
 However, it has been shown that in practice, using different models seem to be produce viable error estimates.~\cite{Sav-JCP-20,PolMadSav-INC-22}
 As these are based on the Taylor expansion in $1/\mu$ finding bounds is not out of question.
 Extending this approach to two different basis set seems feasible.
 We can discover in the plots shown in this paper the stability of the energy estimates from calculations with two different basis sets.
 At this point we would like to stress again that ABC does only correct for the errors due to the singularity of the electron-electron interaction: the errors with the smallest basis sets seen in the example show up already at Hartree-Fock level, and that these need to be estimated first.
}
 \item A regularized expression could prove to be useful.
   \updated{}{It is easily shown that the rational approximant of eq.~\eqref{eq:reg} is larger or equal to the ABC correction.}
    We suspect that a discrepancy between \updated{}{a regularized expression, \textit{e.g.},} PBE-OT and our results could serve as a red flag indicating departure from the large-$\mu$ domain where ABC is valid.
 The He in the VDZ, or the O atom in the VTZ basis set are examples.
 \updated{As for $e_c \le 0$ and $a_3 < 0$, \dots
    the PBE regularized basis-set correction is always an upper bound to our correction.}{}
 An avenue for future exploration is regularization via other means, \textit{e.g.}, by imposing some exact conditions.~\cite{SceSav-Trygve-JPCA-24}
 \item Since ABC does not rely on the Hohenberg-Kohn theorem, it can be applied to excited states, and we showed it only for a few examples.
 The study of excited states might often need the simultaneous treatment of several states~\cite{Lis-CR-18}.
 \item Further examples are needed, particularly regarding the calculation of properties.
 \updated{Up to now, we only have the numerical results with properties when using the model that we make equivalent to a basis-set calculation}{So far, we have numerical results for properties only when using the model that we identify with a basis-set calculation} (see, \textit{e.g.}, ref.~\onlinecite{PolMadSav-INC-22}).
 \item \updated{}{A comprehensive benchmark including atomization energies, reaction barriers, bond-breaking curves, multireference systems, and a wider range of excited states will be essential to fully establish the general applicability and robustness of the ABC correction across chemical space.
 However, the primary goal of the present manuscript is to lay down the theoretical foundations of the ABC methodology and to provide a first proof-of-concept validation on representative, well-understood systems where exact or highly accurate reference data are available. Before undertaking large-scale, systematic benchmarks across broad molecular datasets, we believe it is crucial to first thoroughly investigate other fundamental theoretical and computational aspects of the method.}
\end{enumerate}

\section*{Supplementary Material}

See the supplementary material for additional information regarding computational details and numerical data. Specifically, it includes:
\begin{itemize}
    \item Detailed computational parameters, including selection thresholds for CIPSI calculations and Becke numerical integration grid settings (radial and angular points);
    \item Exact reference total energies for two-, three-, and four-electron Harmonium states, as well as atomic and molecular test systems ($\mathrm{H}^-$, $\mathrm{He}$, $\mathrm{O}$, $\mathrm{O}^+$, and $\mathrm{H}_2$);
    \item Full basis-set series ($X = 2$ to $9$) of Hartree--Fock, variational, perturbative, and extrapolated FCI energies;
    \item Tabulated values of the local model parameter $\mu_B(\mathbf{R})$ and uncorrected on-top pair densities $P_2(\mathbf{R},\mathbf{R})$ as a function of position $R$;
    \item Exponents and angular momenta of the even-tempered Gaussian basis sets used for Harmonium calculations;
    \item Complete numerical data comparing absolute energy errors relative to reference benchmarks across the FCI, CBS, PBE-OT, and ABC methods.
\end{itemize}

\section*{Acknowledgement}

We would like to dedicate this paper to the memory of Axel Becke.
Although his fame is mainly based on the success of his density functional approximations, \updated{he always wanted to understand what he is doing}{he always sought to understand the principles underlying his methods}.
For example, he observed for many years that a mixing between density functional and Hartree-Fock results could improve the results.
However, he did not publish it, until he found the adiabatic connection as a motivation for such a mixing.
The present paper is in this spirit.

We also would like to acknowledge productive discussions with Emmanuel Giner (CNRS and Sorbonne University, Paris).
\updated{}{We are grateful to Kasia Pernal for pointing out ref~\onlinecite{HapTucModGolVeiPer-JPCL-25}.}

This work used the HPC resources from CALMIP (Toulouse) under
allocation 2026-18005.

\bibliography{biblio}
\end{document}